# All Optical Implementation of Multi-Spin Entanglement in a Semiconductor Quantum Well

J. Bao,[1] A. V. Bragas,[1] J. K. Furdyna,[2] and R. Merlin[1]

[1] *FOCUS Center and Department of Physics, The University of Michigan, Ann Arbor, Michigan 48109-1120*

[2] *Department of Physics, University of Notre Dame, Notre Dame, Indiana 46556*

June 27, 2002

We use ultrafast optical pulses and coherent techniques to create spin entangled states of non-interacting electrons bound to donors (at least three) and at least two $Mn^{2+}$ ions in a CdTe quantum well. Our method, relying on the exchange interaction between localized excitons and paramagnetic impurities, can in principle be applied to entangle a large number of spins.

The problem of quantum entanglement has attracted much attention since the early days of quantum mechanics. One of the most intriguing features of this phenomenon is that non-interacting parts of a system can show non-local correlations reflecting interactions that occurred in the past [1]. Although techniques to entangle two particles and, in particular, two photons [2] have been known for some time, it is only recently that many (up to four) - particle entanglements have been demonstrated experimentally [3-5]. Following recent developments in quantum cryptography together with the discovery of quantum algorithms that outperform those of classical computation, research on the foundations of quantum mechanics has now moved to the center of the new field of quantum information [6]. As a result, the question of entanglement has acquired practical significance.

In this work, we propose and demonstrate experimentally a novel method for many-particle entanglement involving Zeeman-split spin states of *non-interacting* paramagnetic impurities ($Mn^{2+}$ and donors) in a CdTe quantum-well (QW). Our approach relies on the exchange interaction between the impurity spin and excitons that are generated and controlled by optical means. As such, our scheme belongs to the category of entanglement mediated by an auxiliary quantum particle (e. g., the center-of-mass phonon in ion-trap quantum computers [5]), as opposed to that resulting from the variation of parameters such as external fields or exchange constants [7]. Because the entangled spins do not interact among themselves unless an exciton is present, our method distinguishes itself from other approaches where the source of entanglement are interactions that cannot be controlled by the experimenter (e. g., excitons in quantum dots [8] and NMR quantum computers [6,9]). Our approach is in principle scalable, thereby holding promise for applications in quantum computation.

Our entanglement procedure draws from an idea put forth by Stühler et al. [10] to explain multiple $Mn^{2+}$ spin-flip Raman scattering (RS) in the same materials system we use to demonstrate the method [11]. The relevant exciton and spin energy levels are shown schematically in Fig. 1. $\vec{S} = (S_x, S_y, S_z)$ is the *total* spin of $N$ impurities and $\mathbf{B}_e$ is an effective magnetic field describing the exchange coupling between the exciton and the impurities [10,12]. Central to our scheme is the requirement that $\mathbf{B}_e$ not be parallel to the external field $\mathbf{B}$, so that the quantization axes in the ground ($x$-axis) and in the optically-excited state ($w$-axis) not be the same, and that $<S_z> = \Delta_S \neq 0$ when the exciton is present. This is made possible by a combination of quantum confinement and spin-orbit coupling leading to heavy-hole ($m_J = \pm 3/2$) states for which the hole-spin is parallel to the QW $z$-axis irrespective of the orientation of $\mathbf{B}$ (this, provided the splitting between the heavy- and the $m_J = \pm 1/2$ light- hole states is large compared with the Zeeman splitting [13]). Since the eigenstates of $S_x$ and $S_w$ are not orthogonal to each other, this opens optical paths, such as the one shown by the red arrows, that can be used to establish Raman coherences between an arbitrary pair of eigenstates of $S_x$ and, independently, of $S_w$. Focusing for the moment on the $S_x$–ladder, we then expect that a single properly tailored optical pulse will be able to create *entangled* superposition states of the form $\Psi = \sum_{k=0}^{2S} C_k e^{ik\Omega_0 t} |S-k\rangle$ with predetermined values for the parameters $C_k$ [14]. Here $|\vec{S}|^2 \Psi = S(S+1)\Psi$, $|S-k\rangle$ is the eigenstate of $S_x$ of eigenvalue $(S-k)$ and $\Omega_0 = g\mu_B B/\hbar$. For $\mathcal{S} \gg 1$, we can map the problem into that of two harmonic oscillators displaced by $\Delta_\mathcal{S}$ with respect to each other, as represented by the parabolas in Fig. 1 [12]. Thus, the Hamiltonian becomes identical to that of a molecule with electrons that couple to a single vibrational mode. The associated Huang-Rhys factor is given by $\Delta_\mathcal{S}^2/2 = \mathcal{S}(1+B^2/B_e^2)^{-1}/2$. This mapping is important, for it allows us to use the vast literature on coherent molecular spectroscopy (particularly time-domain studies [15]) to analyze the experimental results.

The measurements were performed on a multiple QW structure consisting of 100 periods of 58-Å-thick $Cd_{0.997}Mn_{0.003}Te$ wells and 19-Å-thick MnTe barriers grown in the [001] $z$-direction [16,17]. The sample is nominally undoped. Consistent with other reports [10,18], however, our RS experiments reveal the presence of isolated donors in the wells. We used both a cw and a mode-locked Ti-sapphire laser. The latter provides ~ 130 fs pulses at a repetition rate of 82 MHz which were focused to a 400-μm-diameter spot using an average power of 3-4 mW. RS, photoluminescence (PL) and PL excitation spectra were obtained with the cw source using power densities of ~ $10^{-2}$ Wcm$^{-2}$. Data were obtained either in the Voigt ($\mathbf{B} \perp z$) or the Faraday ($\mathbf{B}//z$) geometry with the photon wavevector along $z$. The laser range covers the fundamental gap of the wells, at ~ 1.68 eV, associated with dipole-allowed transitions involving heavy-hole states. Time domain data were obtained using a standard pump-probe setup in the reflection geometry. Changes in the intensity of the reflected probe pulse, $\Delta R$, give a measure of the dependence of the optical constants on the coefficients $C_k$ characterizing the particular entangled states created by the pump. The signature for a Raman coherence involving states $|\mathcal{S}-j\rangle$ and $|\mathcal{S}-l\rangle$ is the observation of oscillations of frequency $|j-l|\Omega_0$. Following the molecular nomenclature, the entangled states with and without excitons will be referred to in the following as "excited-" and "ground-state" coherences. Processes by which ground-state coherences modulate the probe signal are impulsive stimulated RS and its associated imaginary component, nonlinear absorption. Excited state coherences modulate the spectrum of stimulated emission [15].

Resonant Raman and differential reflectivity spectra were obtained using photon energies near the QW gap. The dominant features are *intra-well* excitations involving the $^6S_{5/2}$-multiplet of $Mn^{2+}$ and spin-flips associated with states derived from the conduction band. The field dependence of these transitions is given by the Zeeman coupling $\mu_0 \mathbf{B} \cdot (g_c \sigma + g_{Mn}\mathbf{S})$ where $\sigma$ and $\mathbf{S}$ are, respectively, the electron spin of the donor ($\sigma = \frac{1}{2}$) and the $Mn^{2+}$ ions ($S = 5/2$) with $g_{Mn} \approx 2$ [16]. The parameter $g_c = \tilde{g}_c - n_0 \alpha x \langle \mathbf{S} \cdot \mathbf{B} \rangle / \mu_0 B^2$ is an effective factor that accounts for the exchange interaction of the donor electron with the $Mn^{2+}$–spins. Here $\tilde{g}_c \approx -1.64$ is the bare g–factor, $x$ is the $Mn^{2+}$ concentration and $n_0$ is the density of unit cells; for CdTe, $n_0\alpha \approx 0.22$ eV [16]. Coupling to the $Mn^{2+}$ ions leads to a strongly nonlinear dependence of the Zeeman levels with $B$, characterized by saturation, that can be used to determine the $Mn^{2+}$-concentration as well as to distinguish electron spin-flip from other features (note that the g–factor has approximately the same value for free electrons, electrons bound to donors and electrons in free and bound excitons) [16].

Resonant Raman spectra (not shown) obtained in the Voigt geometry at $B = 7$T show ~ 10 lines at multiples of the $Mn^{2+}$ paramagnetic resonant (PR) frequency and, in addition, a peak at ~ 13 cm$^{-1}$ due to the spin-flip of an electron bound to a donor [18]. The PR-scattering is strongly enhanced when the energy of the scattered photon resonates with the main PL feature at $E_e$ ~ 1.677 eV. Our RS results are in excellent agreement with the work of Stühler et al. [10] on similar samples. As in [10], we find that the position of the PL is red-shifted (by ~ 6 meV at $B = 0$) with respect to the maximum of the PL excitation spectrum, indicating that the relevant intermediate states of the Raman process are localized excitons. Consistent with the model depicted in Fig. 1, the multiple PR-scattering is not observed in the Faraday geometry where the hole quantization axis, **B** and **B**$_e$ are all along $z$. We recall that the peaks involving an odd (even) number of Mn$^{2+}$ spin-flips are observed mainly when the polarizations of the incident and scattered light are perpendicular (parallel) to each other [10]. This agrees with the fact that the corresponding excitations are odd (even) under time-reversal and, hence, that they belong to the antisymmetric $A_2$ (symmetric $A_1$) representation of the $D_{2d}$ point group of the QW (however, a departure from this symmetry is observed at very low temperatures [10]).

Figure 2 shows time-domain data obtained in the Voigt geometry using a circularly polarized pump (which couples to only one of the two $m_J = \pm 3/2$ states) and a probe detection scheme for which only antisymmetric $A_2$ excitations are allowed [19]. We worked at sufficiently low laser fluences so that the normalized differential reflectivity, $\Delta R/R$, depends linearly on the pump and is independent of the probe intensity. The long-lived oscillations are due to the Mn$^{2+}$ PR-transition. At short times, the signal is dominated by electron Zeeman quantum beats, i. e., oscillations showing a field- and temperature-behavior consistent with that of spin-flips of electrons. The corresponding feature in the Fourier transform (FT) spectrum is labeled SF. These results bear a close resemblance to those reported for (Zn,Cd,Mn)Se heterostructures by Crooker et al. [20] who ascribed the rapidly decaying oscillations to free photoexcited electrons. However, close inspection of the data brings out important differences concerning the nature of the oscillations as well as the source of the coherence. As shown in Fig. 3, linear prediction fits [21] reveal that (*i*) the dominant SF feature is actually a doublet at ~ 12 and 13 cm$^{-1}$ (*B* = 7 T), and that there are additional contributions due to (*ii*) a mode labeled PR(*e*) whose frequency depends weakly on the field and, most strikingly, (*iii*) peaks 2SF and 3SF which appear at nearly twice and three times the frequency of SF. These modes do not turn up in the RS spectra but, like the Raman features, their amplitudes exhibit a large enhancement when the central energy of the pulses, $\hbar\omega_C$, is tuned to resonate with the PL associated with localized states. Because the spectrum of localized excitations is very susceptible to Mn$^{2+}$ density fluctuations, and also depends on the positions of the localization centers in the QW, we believe that the fast decay of all but the PR-mode is caused by inhomogeneous broadening.

For various reasons, including the important fact that they do not come out in the RS data, we assign all but the PR-feature in Fig. 2 to spin-flip *excited-state* coherences. Explicitly, we ascribe (*i*) the low- (high-) frequency component of the SF doublet to the spin-flip of the electron that belongs to the donor (exciton), (*ii*) the PR(*e*) mode to the Mn$^{2+}$ transition in the excited state and (*iii*) the 2SF and 3SF peaks to multiple spin-flips of donors. The observation of SF-overtones is central to our claim of multi-spin entanglement. Since the electron spin is $\sigma = \frac{1}{2}$, the presence of these modes indicates the establishment of Raman coherences and, hence, entanglement involving at least three donor impurities. These assignments are substantiated in the following.

Consider the heavy-hole component of the exchange interaction $V_{HH} = (\beta/3)\sum_i \delta(\mathbf{r}-\mathbf{r}_i)\mathbf{S}_i \cdot \mathbf{J}$ between a QW exciton and $Mn^{2+}$ ions at sites $\mathbf{r}_i$ where $\beta$ is a constant, $\mathbf{r}$ is the coordinate and $\mathbf{J} // z$ is the pseudo total angular momentum of the heavy-hole ($|\mathbf{J}| = 3/2$) [16]. As mentioned earlier, the effect of $V_{HH}$ on the ions can be expressed in terms of an effective field $\mathbf{B}_e(\mathbf{r}) = (\beta/3\mu_0 g_{Mn})|\Psi_{HH}(\mathbf{r})|^2 \mathbf{J}$ where $\Psi_{HH}$ is the hole wavefunction [10]. Hence, in the presence of the exciton the paramagnetic transition energy evolves from $\mu_0 g_{Mn} B_e$ at $B = 0$ to $\mu_0 g_{Mn} B$ for $B \gg B_e$. The latter is consistent with the asymptotic behavior and the basis for assigning the PR(*e*) peak to the $Mn^{2+}$ spin-flip transition in the excited state. From the PR(*e*) frequency at zero field and using $n_0\beta \approx 0.88$ eV for CdTe [16], we obtain the very reasonable estimate of ~ 40 Å for the hole localization length (the bulk exciton radius in CdTe is 50Å) leading on average to 2.5 $Mn^{2+}$ ions per hole. From this number, the Huang-Rhys factor we infer is ~ 1.06 ($B = 7$T) which is consistent with the observation of ~ 10 overtones in the Raman spectrum. Albeit indirect, the combination of the RS and coherent results gives strong evidence for exciton-mediated entanglement involving at least two $Mn^{2+}$ ions. Since the total spin of a two-ion system is $\mathcal{S} = 5$, its spectrum is well described by the vibrational analog. Given that impulsive excitation of a harmonic mode does not give high-order harmonics [15], this provides an explanation as to why $Mn^{2+}$-overtones are not seen in the time-domain data (this does not apply to the $\mathcal{S} = 1.5$ three-donor case which, with only four levels, is strongly anharmonic). We further note that, since ground state amplitudes are significantly smaller than those for the excited state at weak pump intensities [15], the vibrational model accounts also for the fact that the signal is dominated by excited state coherences.

Our assignment of the SF doublet, 2SF and 3SF, as due to the spin-flips of bound electrons is supported, first, by the observation that these features resonate at $\hbar\omega_C \approx E_e$ and, second, by their dependence on temperature and field which show excellent agreement with theoretical predictions. The curves in Fig. 3 are fits using the standard Brillouin function to account for $<\mathbf{S}.\mathbf{B}>$ in the expression for the effective electron parameter $g_c$. The resonant behavior is a clear indication that the relevant states are localized. This, and the linear dependence of the signal with the pump intensity are consistent with the donor interpretation for 2SF and 3SF, since linearity excludes the possibility that the overtones could be due to multiple spin-flips of bound excitons. Additional proof is given by the fact that 2SF is usually much weaker than 3SF; see Fig. 2. This supports our assignment since double flip excitations transform like $A_1$ and, therefore, are nominally forbidden in the geometry we used (all other modes, being single or triple flips, belong to the $A_2$ representation). Based on the value of the frequencies at large fields, the fundamental mode from which 2SF and 3SF derive is ascribed to the lower-frequency component of the SF doublet. This mode and the overtones exhibit a similar $\hbar\omega_C$–behavior, different from that of the higher-frequency component associated with the spin-flip of the electron in the bound exciton (see Fig. 1). The occurrence of the SF-doublet is attributed to exchange effects in the excited-state. From measured the splitting, we obtain an upper limit of ~ 600 μeV for the electron-hole exchange that is consistent with the value 270 μeV from the literature [22]. This estimate ignores electron-electron exchange which is large for exciton-donor complexes [23] and may provide an additional longer-range mechanism for donor entanglement.

In conclusion, our results confirm that there is a system of localized excitons coupled to paramagnetic impurities in a CdTe QW that is well described by the level structure of Fig. 1. We have shown that such a system can be optically excited to generate many-spin Raman coherences and, thus, entanglements involving multiple impurities. Since the exciton localization centers (due to, e. g., surface roughness) and the impurities need not be at the same sites, our scheme can in principle be applied to entangle an arbitrary number of spins by using different excitons to address different sets of impurities, making it potentially useful for quantum computing applications.

Acknowledgment is made to the donors of The Petroleum Research Fund, administered by the ACS, for partial support of this research. Work also supported by the NSF under Grants No. PHY 0114336 and No. DMR 0072897, by the AFOSR under contract F49620-00-1-0328 through the MURI program and by the DARPA-SpinS program. One of us (AVB) acknowledges partial support from CONICET, Argentina.

geometry employed by these authors in their differential Faraday rotation experiments allows for $A_2$- but not $A_1$-symmetry excitations.

**FIGURE CAPTIONS**

FIG. 1 (color) – Generic level structure of a QW system of electrons bound to paramagnetic impurities (green circles) coupled to a localized exciton of energy $E_e$. Large (small) blue circles represent the hole (electron) in the exciton. Optical transitions are depicted by red arrows and spins by arrows attached to the circles. $\mathcal{S}$ is the total spin quantum number. The diagram shows two exciton states that differ in the orientation of the electron spin. Parabolas represent the vibrational analog in the limit $\mathcal{S} \gg 1$.

FIG. 2 – Differential reflectivity data at $\hbar\omega_C = 1.687$ eV (circles). The reflected beam was not spectrally analyzed. Curves are fits using the linear prediction method [21]. Parameters gained from the fits were used to generate the FT spectra. The dominant SF peak is a doublet associated with electron spin-flips. Its lower-frequency component, exhibiting overtones at twice (2SF) and three times (3SF) the fundamental frequency, is due to electrons bound to donors. The narrow PR and PR($e$) are ground- and excited-state $Mn^{2+}$ spin-flip transitions.

FIG. 3 (color) – Frequency vs. magnetic field at $\hbar\omega_C = 1.687$ eV. Black (red) symbols represent spin-flips of donor (exciton) electrons. Associated curves are fits using the Brillouin function [16]. $Mn^{2+}$ transitions are shown in blue. Lines are linear fits to the data.

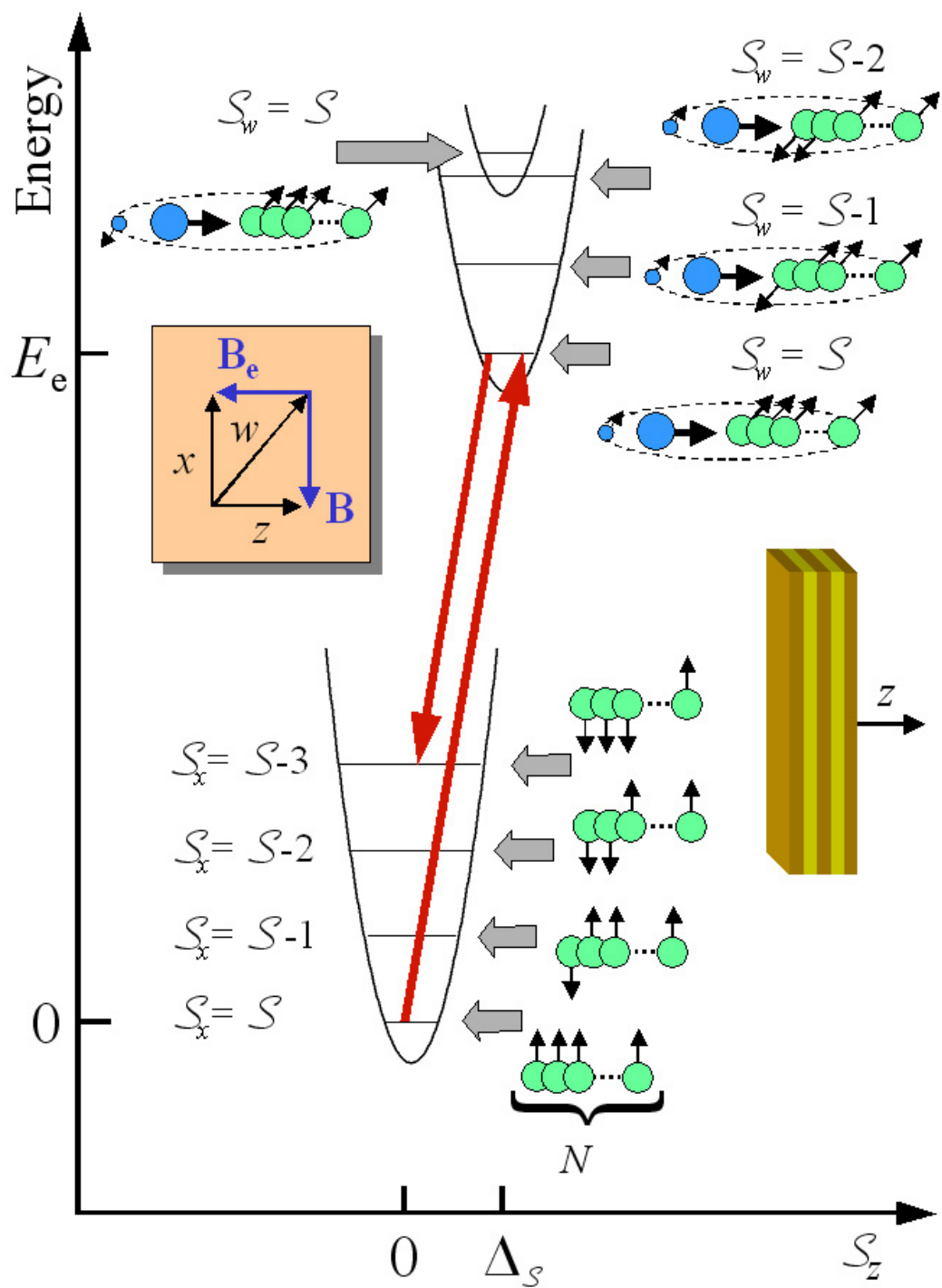

Figure 1

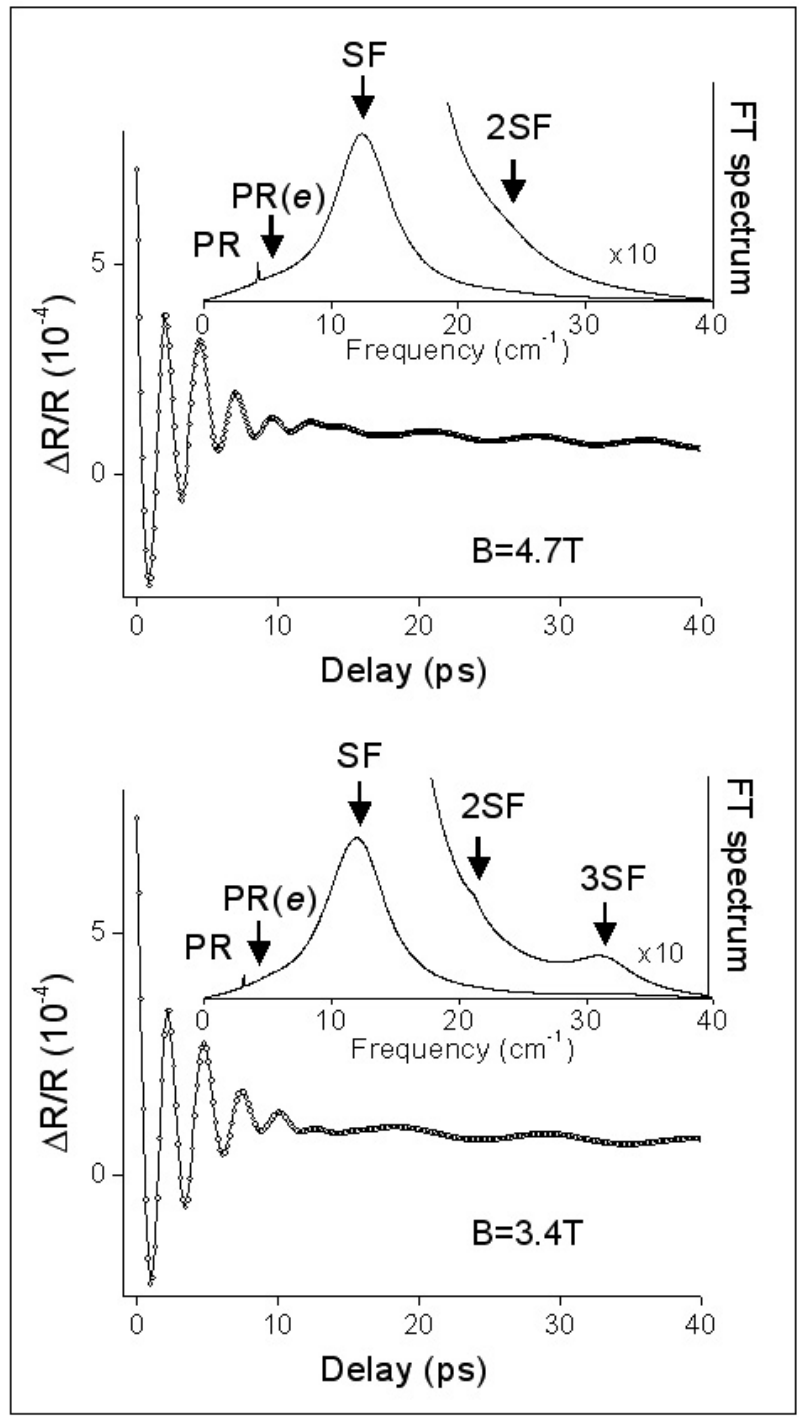

Figure 2

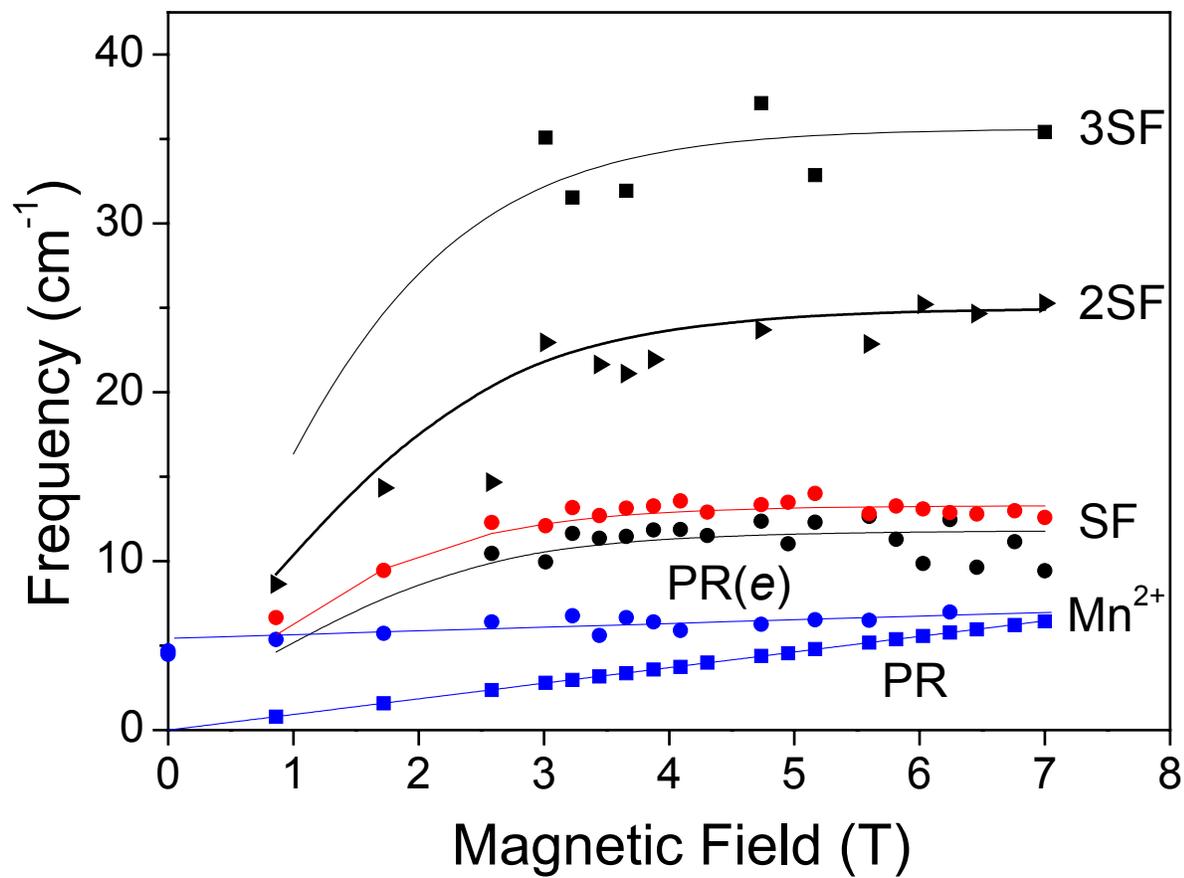

Figure 3